\documentstyle[fleqn,12pt]{article}
    \title{Zeroing in on More Photons and Gluons}
    \author{Robert W.Brown$^{\dag}$, Mary E.
    Convery $^{\dag}$ and Mark A. Samuel$^{\ddag\ast}$}
    \date{}
    
\begin{document}
    \maketitle
    {\centerline{\it $^{\dag}$Department of Physics, Case Western
    Reserve University, Cleveland, OH 44106}}
    {\centerline{\it $^{\ddag}$Department of Physics, Oklahoma State  
    University, Stillwater, OK 74078}}
    {\centerline{\it $^{\ast}$Stanford Linear Accelerator  Center, Stanford
    University, Stanford, CA 94309}}
    \vskip2ex
    
    \begin{abstract}
    We discuss radiation zeros that are found in gauge tree amplitudes for
    processes involving multi-photon emission.
    Previous results are clarified by examples and by further elaboration. 
    The conditions under which such amplitude zeros occur
    are identical in form to those for the single-photon zeros, and 
    all radiated photons must travel parallel to each other.  Any other
    neutral particle likewise must be massless (e.g. gluon) and travel in that common direction.
    The relevance to questions like gluon jet identification
    and computational checks is considered.  We use examples to show how
    certain multi-photon amplitudes evade the zeros, and to demonstrate the
    connection to a more general result, the decoupling of an 
    external electromagnetic plane wave in the ``null zone".  Brief comments
    are made about zeros associated with other gauge-boson emission. 
    \end{abstract}
    \vfill
    \newpage
    \setcounter{equation}{0}

    \section{Introduction }
    
    It is now more than a dozen years since radiation amplitude
    zeros were first discovered~\cite{MSS} in the process $u\overline{d} \rightarrow W^+ \gamma$ ~\cite{BSM}.
    Subsequently, it was shown~\cite{BB,SAM,BKB} that these can arise more 
    generally, originating as the
    destructive interference of radiation patterns in gauge-theory tree amplitudes 
    for massless gauge-boson emission.  This is therefore a property of gauge 
    theories;  anomalous electromagnetic moments, for example, would spoil the 
    perfect cancellations and such anomalies are forbidden in gauge couplings.  
    For a specific analysis of the effect of $W$ anomalous moments in the $u\overline{d} \rightarrow W^+ \gamma$ reaction, see~\cite{SSSS}. 
    Of course, anomalous moments come up in higher-order corrections, and indeed radiation 
    zeros do not appear beyond the tree approximation in any theory.
    
    Can we observe these zeros experimentally?  Since we must be able to use the tree approximation, the couplings have to be small for the process considered.  The equations that determine where the zeros are - essentially these are just 
    the demand that the ratio of coupling to light-cone energy be the same for all 
    particles -   specifically require that all couplings have the same sign 
    (electric charges, for photon emission).  These two constraints, weak couplings 
    and same-sign charges, have very much limited the number of reactions in which 
    a RdipS would be found.   In high-energy quark reactions where gluon emission
    can be described perturbatively, the color charges, unfortunately, are 
    ultimately averaged or summed over.  The benchmark $e^+e^-$ reactions violate the 
    like-sign condition.  Even if certain hadronic reactions involved electric 
    charges of only one sign, and in some limit could be approximated by 
    tree amplitudes, hadrons with spin are composite particles with anomalous 
    moments ($g\neq 2$).  And small cross sections are perforce hard to measure!
    
    Despite the difficulty in ``measuring" zeros in experiments, the 
    sensitivity of the basic quark amplitude $u\overline{d}\rightarrow W^{+}
    \gamma$ to the $W$-boson magnetic moment has attracted much interest in the hope
    that this important parameter could be measured in proton-antiproton 
    colliders~\cite{Mont}.
    The radiation zero is present only for a magnetic moment value 
    corresponding to $g=2$ as
    predicted by gauge theory.  The extent to which there is a pronounced dip will 
    make it possible to put limits on the $W$ magnetic moment (see Sec. V);  
    experimenters can also consider the crossed channel reaction, radiative 
    $W$-decay, whose zero shows up in the energy distributions~\cite{GM}.  A 
    basic $e^{-}e^{-}\rightarrow e^{-}e^{-}\gamma$ radiation zero is irrelevant to 
    present accelerators, but there is the possibility that HERA experiments may 
    probe a radiation amplitude zero in electron-quark bremsstrahlung and allow a 
    direct measurement of the quark charge~\cite{Bilc,Cout,RLS}.
    
    The very welcome progress in accelerator experimentation brings with it a challenge.  We can anticipate more detailed information, not only in the way of more single-photon events, but also in the variety of final states measured.  Are there electromgnetic radiation zeros in reactions with more final photons,
    such as two-photon exclusive reactions?  What about the associated production in the QCD perturbative regime of another massless gauge boson of great
    interest, the gluon, such as in $u\overline{d}\rightarrow W^{+}
    \gamma g$?  
    
    The answer to both is that, yes, in general the radiation zeros survive the addition
    of more neutral, massless particles.  If a given reaction has an
    electromagnetic radiation zero in its tree amplitude, then the tree reaction with additional photons and gluons produced in the final state will too, occurring when these additional particles travel parallel to the original photon,
    sharing its original energy.  This answer is given to us already in Ref.~\cite{BKB}.  
    
    In the present paper, we revisit this question, in view of the experimental change of scenery and the fact that the higher-order zeros are not so well
    known.  The purpose is to call attention to our previous results on 
    multi-photon/gluon
    electromagnetic radiation zeros, and to try to combine them into a self-contained and clear picture through detailed examples.  
    The examples are laid out in Sec. II.  In the next two sections, we use the
    examples to illustrate the
    survival theorem ~\cite{BKB} for neutral, massless particles, and the decoupling
    theorem~\cite{BK,BK2} for an external electromagnetic plane wave field.
    Finally, we consider the relevance of the multi-gauge-boson emission zeros as they pertain to the new generation of electron-proton and proton-antiproton colliding beams.

    \setcounter{equation}{0}
    \section{Examples}
    
    We wish to present several tree amplitudes for the emission/absorption of multiple gauge bosons.  The first intention is to exhibit their
 
    zero structure, and also to show a simple counter-example, in which there are multiple photons but neither a physical nor unphysical radiation zero.  Second, we focus on a reaction relevant to experiment. 
    
    \subsection{Scalar particles and photons}

    Consider the radiative process where a scalar particle decays in
    lowest order through a single-vertex scalar interaction into $n-1$ other scalar 
    particles plus one photon.  Denoting the electric charges by $Q_i$, the 
    reaction is
    \begin{eqnarray*}
    Q_1 & \rightarrow & Q_2\:+\ldots +\:Q_n\:+\:\gamma (q)
    \end{eqnarray*}
    The diagrams of the tree amplitude are illustrated in Fig. 1 and we 
    can write it as 
    \begin{equation}
    M=\sum_{i=1}^n\left( {Q_i\over p_i\cdot q}-{Q_j\over p_j \cdot q}\right)
    g\delta_i p_i\cdot\epsilon 
    \label{eq:onephot}\end{equation}
    where $j$ is fixed and can take any $i$ value, and $\delta_i=-1 (+1)$ for incoming (outgoing)
    particles.   
    
    The fact that we could rearrange (factorize) (\ref{eq:onephot}) as shown~\cite{GHL,BB,SAM,BKB} 
    is due to the presence of the zero.  It is evident that the amplitude vanishes 
    in the null zone defined by the $n-1$ equations [these actually reduce
    to $n-2$ independent ones by charge and momentum conservation]
    \begin{eqnarray}
    {Q_i\over p_i\cdot q}&=&{Q_j\over p_j\cdot q},\;\;\;\mbox { all  
    $i,\:j$ } \label{eq:same}
    \end{eqnarray}
    That is, the kinematical conditions for the null radiation zone are that all 
    particles must have the same charge to light-front-energy ratio.  The ratios 
    are recognized as the factors arising from the attachment of a photon to the 
    various external lines, with the remarkable feature that these conditions also 
    suffice to cancel out internal-line attachments in the tree amplitudes.  (One 
    sees immediately why closed-loop, higher-order amplitudes will not be nullified:
    Integrated internal loop momenta are certainly not fixed.)  
    
    We digress for the moment.  Recall that the radiation amplitude zero is not 
    spoiled by photons attached to 
    internal tree lines.  For example, if we look at a tree source graph with one
    internal line, the photon attachments can be rearranged into a sum over two
    vertex terms.  The vertex terms can themselves be rearranged as in
    (\ref{eq:onephot}). 
    In particular, consider the process
    \begin{eqnarray*}
    Q_1\rightarrow Q_2 +(Q_{int}\rightarrow Q_3+\ldots +Q_n) +\gamma (k)
    \end{eqnarray*}
    where $Q_{int}=Q_1-Q_2$ represents a virtual particle of mass $m$.  The
    total amplitude for this process is obtained by attaching the photon in
    all possible ways to the external lines and also the internal line (see
    Fig. 2).  Using the radiation decomposition identity~\cite{BB,BKB} on the term 
    with photon
    emission from the internal line ($p'=p-q$)
    \begin{equation}
    {1\over p'^2-m^2}Q(p'+p)\cdot\epsilon{1\over p^2-m^2}
    ={1\over p'^2-m^2}{Q\over p' \cdot q} p' \cdot\epsilon
    -p\cdot\epsilon{Q\over p\cdot q}{1\over p^2-m^2}
    \end{equation}
    this amplitude can be written as two clusters corresponding to
    corrections to the two source vertices
    \begin{equation}\begin{array}{rl}
    M= & -i\epsilon^*\cdot\left\{\left[ {Q_1p_1\over p_1\cdot q}
    -{Q_2p_2\over p_2\cdot q} -{Q_{int}(p_1-p_2)\over (p_1-p_2)\cdot q}\right]
    {1\over (p_1-p_2-q)^2-m^2}\right. \\
    & -\left.\left[\sum_{i=3}^n {Q_ip_i\over p_i\cdot q} -{Q_{int}(p_1-p_2)\over
    (p_1-p_2)\cdot q}\right]{1\over (p_1-p_2)^2-m^2}\right\}
    \end{array}\end{equation}
    Since the quantity in each square bracket vanishes under the zero
    conditions, we see that the zeros persist at the same location in phase space,
    independent of the mass of the internal particle.
    In its clustered form this example will help the reader follow the more general discussion given 
    in Sec. III.
    
    Now add another photon. The process we consider is
    \begin{eqnarray*}
    Q_1 & \rightarrow & Q_2\:+\ldots +\:Q_n\:+\:\gamma (q_1)\:+\:\gamma (q_2).
    \end{eqnarray*}
    Although we still restrict ourselves to spinless charges with scalar
    self-interactions, the subsequent discussion will make it clear how spin
    and gauge interactions can be incorporated.  We again consider only a
    single n-scalar vertex.  
    
    It is not hard to write down the lowest-order
    tree amplitude such that the zero is manifest.
    We can rearrange the sum of diagrams, using experience gained from our
    previous factorization study to rewrite some of the terms,
    \begin{equation}
    \begin{array}{rcl}
    M & = &  g\sum_{i=1}^n {\delta_i Q_i\over p_i\cdot q_1} p_i\cdot\epsilon_1^*
    \sum_{k=1}^n \left( {Q_k\over p_k\cdot q_2} -{Q_j\over p_j\cdot q_2}\right)
    (\delta_k p_k+\delta_{ik}q_1)\cdot\epsilon_2^* \\
    & & +g\sum_{i=1}^n \left( {Q_i\over p_i\cdot (q_1+q_2)} -{Q_j\over p_j\cdot
    (q_1+q_2)}\right) \delta_i Q_i (-\epsilon_1^*\cdot\epsilon_2^*) \\
    & & +q_1\cdot q_2 g\sum_{i=1}^n Q_i^2 \left[ {1\over \delta_i p_i\cdot
    (q_1+q_2)+q_1\cdot q_2}\, {1\over p_i\cdot q_2}\, {1\over (\delta_i p_i
    +q_1)\cdot q_2}\, p_i\cdot\epsilon_1^*\, (\delta_i p_i+q_1)\cdot\epsilon_2^*
    \right. \\
    & & \quad -{1\over p_i\cdot q_1}\,{1\over p_i\cdot q_2}\,{1\over (\delta_i
    p_i+q_1)\cdot q_2}\,\delta_ip_i\cdot\epsilon_1^*\,(\delta_ip_i+q_1)\cdot
    \epsilon_2^* \\
    & & \left.\quad +{1\over\delta_ip_i\cdot (q_1+q_2)}\,{1\over\delta_ip_i\cdot
    (q_1+q_2)+q_1\cdot q_2}\,\epsilon_1^*\cdot\epsilon_2^*\right] \\
    & & +g\sum_{i=1}^n {Q_i^2\over\delta_i\, p_i\cdot (q_1+q_2)+q_1\cdot q_2}
    {1\over p_i\cdot q_2} (q_2\cdot\epsilon_1^* p_i\cdot\epsilon_2^*
    -q_1\cdot\epsilon_2^*\, p_i\cdot\epsilon_1^*) 
    \end{array}\label{eq:amp}
    \end{equation}
    
    On the face of it, one might assume that we need three different sets of
    conditions for (\ref{eq:amp}) to vanish.  The first set is the null zone
    conditions for $q_1$
    \begin{eqnarray}
    \frac{Q_i}{p_i \cdot q_1}&=&\frac{Q_j}{p_j \cdot q_1},\;\;\;\mbox{ all
    $i,\:j$  }\label{eq:same2}  
    \end{eqnarray}
    The second set is the analogous conditions for $q_2$
    \begin{eqnarray}
    \frac{Q_i}{p_i \cdot q_2}&=&\frac{Q_j}{p_j \cdot q_2},\;\;\;\mbox{ all
    $i,\:j$ } \label{eq:same3}
    \end{eqnarray}
    And the third is that the two photons must be parallel (their momenta must
    be proportional to the same null vector, $q_1,\;q_2 
    \propto n$).
    
    Actually, one set of null zone conditions is all we need.  From the conditions (\ref{eq:same2}), for example, taken alone, it follows that the 
    second photon, with its zero charge, must be {\bf massless and parallel} to the first
    photon, and therefore the set (\ref{eq:same3}) follows as well.  We refer the
    reader to Sec. III and a related discussion in \cite{BKB}.  
    {\bf Equation (\ref{eq:same2}) is therefore sufficient}.
    
    \subsection{Photons and Gluons}
    
    Next we look at an example very much relevant to experiment.  This will serve to introduce spin, another massless neutral particle, and a parton reaction to which we return later in the paper.  Consider the radiative decay process where quark-antiquak annihilation leads to a $W$-boson plus a gluon plus a photon,
 
    \begin{eqnarray*}
    u \overline{d} & \rightarrow & W^+\:+\:g\:+\:\gamma
    \end{eqnarray*}
    The eight diagrams of its tree amplitude are indicated in Fig. 3.
    Drawing again on our previous experience\cite{BKB}, we can collapse the results 
    into the following form  
    \begin{equation}\begin{array}{rl}
    M & =\, ieG^{ud}_{V-A} \bar{v} (p_2) \left\{ 
    {Q_{2,color}\over 2p_2\cdot k}\left[
    \sum_{i=1}^3\left(
    {Q_i\over p_i\cdot q}-{Q_j\over p_j\cdot q}\right)\,\delta_i p_i\cdot
    \epsilon_q\,(\not k\not\epsilon_k+2p_2\cdot\epsilon_k)
    \not\epsilon_3 \right.\right. \\
    & +\left({Q_1\over p_1\cdot q}-{Q_3\over p_3\cdot q}
    \right) \, (\not k\not\epsilon_k\not\epsilon_q\,
    q\cdot\epsilon_3 -2\not q\,\epsilon_q\cdot\epsilon_3\, p_2\cdot\epsilon_k 
    +2\not\epsilon_q\, q\cdot\epsilon_3\, p_2\cdot\epsilon_k) \\
    & +\left.\left({Q_1\over p_1\cdot q}-{Q_2\over p_2\cdot q}
    \right)\,\not q\not\epsilon_q\, p_2\cdot\epsilon_k
    \not\epsilon_3\,\right] \\
    & -{Q_{1,color}\over 2p_1\cdot k}\left[ \sum_{i=1}^3\left(
    {Q_i\over p_i\cdot q}-{Q_j\over p_j\cdot q}\right)\,\delta_i p_i\cdot
    \epsilon_q\not\epsilon_3\, (\not\epsilon_k 
    \not k+2p_1\cdot\epsilon_k) \right. \\
    & +\left({Q_2\over p_2\cdot q}-{Q_3\over p_3\cdot q}
    \right)\, (\not\epsilon_q\not\epsilon_k\not k\,
    q\cdot\epsilon_3 
    -2\not q\,\epsilon_q\cdot\epsilon_3\, p_1\cdot\epsilon_k
    +2\not\epsilon_q\, q\cdot\epsilon_3\, p_1\cdot\epsilon_k) \\
    & +\left.\left({Q_2\over p_2\cdot q}-{Q_1\over p_1\cdot q}
    \right)\not\epsilon_3\, p_1\cdot
    \epsilon_k\,\not\epsilon_q\not q \right] \\
    & +\left({Q_1Q_{1,color}\over p_1\cdot (q+k)-q\cdot k}-
    {Q_2Q_{2,color}\over p_2\cdot (q+k) -q\cdot k}
    \right)\, \epsilon_q\cdot\epsilon_k\not\epsilon_3 \\  
    & +\mbox{ terms with factors $\not k\not q$, $q\cdot k$, $k\cdot\epsilon_q$, or
    $q\cdot\epsilon_k$ } \left. \right\} \, (1-\gamma_5)\, u(p_1)
    \end{array}\label{eq:gluon}\end{equation} 
    where $j$ is any fixed number $j\in [1,3]$, and $\epsilon_3=\epsilon^*(p_3)$, 
    etc. $Q_{i,color}$ refers to the SU(3) Clebsch-Gordan coefficients; in this
    case $Q_{1,color}=Q_{2,color}$. Notice that we have arranged the expression to
    show $u$-$\bar{d}$ crossing symmetry. 
    
    First we consider the electromagnetic radiation zero.
    It is evident that the amplitude (\ref{eq:gluon}) vanishes in 
    the photon null zone defined by
    \begin{equation}
    \frac{Q_i}{p_i \cdot q}\:=\:\frac{Q_j}{p_j \cdot q},\;\;\;\mbox{ all
    $i,\:j$ }\label{eq:same4}
    \end{equation}
    As before, this is all we need; (\ref{eq:same4}) forces $k\propto q$ and in turn this implies 
    \begin{equation}
    \frac{Q_i}{p_i \cdot k}\:=\:\frac{Q_j}{p_j \cdot k},\;\;\;\mbox{ all
    $i,\:j$ }
    \end{equation}
    
    Second, we consider the chromodynamic radiation zero. We are reminded that there are zeros associated with any gauge
    group when the corresponding massless gauge bosons are emitted~\cite{BKB,GHL}.  In this
    reaction, we can think of two ways its tree amplitude can vanish:
    electromagnetic interference and chromodynamic interference.
    Instead of thinking of the gluon as just another particle ( electrically neutral) produced along with the photon, let us consider it as the radiation
    due to the color charges (some of which are zero; indeed, the photon is now
    the ``neutral" massless co-produced particle that must be parallel to
    the gluon).
    The analogous zeros for massless gluon radiation depend on the color charges.
    The color null zone is defined by
    \begin{equation}
    \frac{Q_{i,color}}{p_i \cdot k}\:=\:\frac{Q_{j,color}}{p_j \cdot k},\;\;\;
    \mbox{ all $i,\:j$ }
    \end{equation}
    But these demand that the colorless $W$-boson be massless (which it is
    not), and that all particles
    be parallel, a singularly uninteresting limit.  One can show, however, that
    the amplitude (\ref{eq:gluon}) does have this unphysical zero.
    
    \subsection{Counterexamples:  Compton}
    
    A question about Compton amplitudes leaps to mind when radiation
    zeros are studied. The null zone for photon-electron elastic scattering,
    for example, is easily seen to be the forward zero-momentum-transfer
    limit.  But it is well-known that the forward amplitude does not
    vanish.  Why is there no
    amplitude zero in this physical limit?
    
    This amplitude can be rearranged as 
    \begin{equation}
    \begin{array}{rl}
    M = i \bar{u}(p_2) & \left[ \left( {Q_1\over p_1\cdot q_1}Q_2  
    -{Q_2\over p_2\cdot q_1}Q_1\right)
    \,\not\epsilon_2^* (\not\epsilon_1\not{q}_1
    -2p_1\cdot\epsilon_1) \right. \\
    & \left. -{Q_1Q_2\over p_2\cdot q_1} (\not\epsilon_1\, q_1\cdot\epsilon_2^* 
    +\not\epsilon_2^* \, q_2\cdot\epsilon_1) \right] u(p_1) \\ 
    & +i{Q_1Q_2\over p_2\cdot q_1}\,\bar{u}(p_2)\not{q}_1\, u(p_2) \, 
    \epsilon_1\cdot\epsilon_2^*
    \end{array}\label{eq:Comp1}\end{equation}
    where the last term does not vanish under the conditions in
    (\ref{eq:same2}).  (Recall that they force $q_1\propto q_2$ so that  
    $q_1\cdot\epsilon_2 =0$, etc.)
    
    And this is not an electron spin effect; the forward amplitude is
    nonzero for photon-boson scattering as well.
    The amplitude for Compton scattering of scalar particles has a similar
    term (now arising from the seagull graph) which is not zero in the null
    zone 
    \begin{equation}\begin{array}{rl}
    M = & 2i\left( {Q_2\over p_2\cdot q_1} \,Q_1 -{Q_1\over
    p_1\cdot q_1}\, Q_2\right)\, p_1\cdot\epsilon_1\, p_2\cdot\epsilon_2^* \\
    & -2i{Q_1Q_2\over p_2\cdot q_1} \left[ (p_1-q_2)\cdot\epsilon_1\,
    q_1\cdot\epsilon_2^* -q_2\cdot\epsilon_1\, p_2\cdot\epsilon_2^*\right] \\
    & +2iQ_1Q_2\,\epsilon_1\cdot\epsilon_2^* 
    \end{array}\label{eq:Comp2}\end{equation}

    The reason the Compton amplitudes are not null in the null zone lies in
    the forward limit where there is no momentum transfer.  We turn our
    attention to the general proof in order to understand, among other
    things, this exceptional
    case.
    
    \setcounter{equation}{0}
    \section{ The General Result}
    
    We can understand where there are multi-photon zeros in 
    gauge
    theoretic tree amplitudes by appealing to a general radiation
    interference theorem for single-photon zeros and certain neutral
    particle lemmas associated with it~\cite{BKB}.  In this section, we
    revisit the proof of those lemmas to show two things:  First, how
    the examples of the previous section fit into the arguments, with
    the neutral particles identified as additional photons.  Second, how
    the proof is readily generalizable to multi-boson zeros for the emission of
    other massless gauge bosons.  This opportunity lets us reference
    also a larger, unpublished version~\cite{BKB1} of the previous work. 
    
    The presence of a radiation zero for single-photon emission 
    means that it possible to rewrite the tree amplitude in a factorized 
    form, really, a sum of factored terms in one-to-one correspondence 
    to the set of independent conditions, such as those in 
    (\ref{eq:same}).  A radiation representation has been found 
    in~\cite{BKB} where the 
    formulas are organized according to the original vertices in the 
    RsourceS graphs to which the photons would be attached.  The 
    radiation amplitude can be written as a sum of vertex attachments 
    with coefficients related to the rest of the original source graph.  
    Each vertex term $V$ has a radiation representation of the form
    \begin{equation}
    M_{\gamma}(V)\: =\:\sum_{i=1}^{n}\delta_i p_i \cdot q
    \Delta_{ij}(Q)\Delta_{ik}(\delta J)\label{equa31}
    \end{equation}
    with any fixed choice for $j$ and $k$, and the definition
    \begin{equation}
    \Delta_{ij}(x)={x_i\over p_i\cdot q}-{x_j\over p_j\cdot q}
    \end{equation}
    It is supposed that there are $n$ internal and external legs on the 
    vertex, and $J_{i}$ is the product of the photon-emission current for 
    the $i^{th}$ leg and the remaining factors of the original vertex 
    amplitude.  For other gauge groups, the charges $Q_i$ refer to the
    Clebsch-Gordan coefficients coupling an incoming particle to an outgoing
    particle through the gauge boson vertex.  See Secs. VI and X of ~\cite{BKB} for more detail.
    
    We can use the above development for a multi-photon 
    argument, 
    by considering one of the source's external legs to correspond to 
    another photon.  (Although neutral internal lines are not of interest 
    to us, they do not spoil radiation zeros anyway.)   It might appear 
    that there is no zero if one of the original external particles $r$ has 
    zero charge, $Q_r = 0$.  One term in a $\Delta$ factor in 
    (\ref{equa31}) is eliminated and hence that factor will not vanish in 
    the null zone.  But this is not the whole story.
    
    Looking at the terms in the $\Delta$ factors in (\ref{equa31}), 
    we still get zero if, in addition to  $\Delta_{ij}(Q)\:=\:0$ 
    (for $j,k\neq r$) we have $p_r\cdot q = 0$ and $J_r = 0$ in the null zone.  
    So the neutral particle must be massless (which is fine for photons!) and 
    travel parallel to the photon (which are just the conditions 
    derived~\cite{BKB,BKB1} from 
    the zero-charge limit, $Q_r \rightarrow 0$, of the null zone 
    equations, forcing $p_r\cdot q 
    \rightarrow 0$).  Furthermore $J_r$ can vanish when $p_r 
    \propto\:q$ for a massless vector neutral 
    particle $r$ if it is coupled to a conserved current (which is also fine 
    for photons!) in a ``nonforward" direction (explained below).  To 
    understand this vanishing for a given photon attachment, let us go 
    over the various current contributions for $p_r 
    \propto\:q$.  The convection current 
    $p_r \cdot \epsilon$ is clearly zero, as is the contact current which 
    involves
    the contraction $p_r^{\alpha}\:\omega_{\alpha \beta}$ where $\omega_{\alpha \beta}\:=\:q_{\alpha}\:\epsilon_{\beta}\:-\:\epsilon_{\alpha}\:q_{\beta}$. The vector 
    spin current $\omega_{\alpha \beta} 
    \eta_r^{\beta}\:=\:q_{\alpha}\:\epsilon \cdot
    \eta_r$ contains the factor $q_{\alpha}$ (which is to be contracted into
    the conserved current vertex source of the vector particle).  This factor is proportional 
    to the momentum transferred to the vertex, $\Delta p\:=\:p_r\:\pm\:q$ for 
    photon emission from a particle in the 
    final/initial state. Thus, if the momentum transfer $\Delta p$ is nonzero 
    (``nonforward scattering"), the vector spin current contribution vanishes by current conservation.  When the momentum transfer is zero, however, we no
    longer have any proportionality relation, the
    vector spin current contribution is not zero, and neither is the amplitude.
    
    The point is that if the null zone corresponds to forward
    scattering of massless vector particles (like photons) then the amplitude
    is not null.  The last terms in the Dirac and scalar Compton amplitudes,
    (\ref{eq:Comp1}) and (\ref{eq:Comp2}), respectively, do not vanish under
    the null zone conditions, and exemplify the vector currents of which we

    The lemmas in the references~\cite{BKB,BKB1} tell us that an 
    arbitrary number of neutral external particles can coexist with a 
    radiation zero, as long as they are all massless and all travel parallel 
    to the photon, and hence to each other.  We can take the special case 
    of their all being photons, each certainly coupled to a conserved 
    current, and we merely need to avoid forward scattering limits 
    where the picture is of a subset of initial photons turning into a 
    subset of final photons without a change in the overall momentum of 
    the photon ``pack". Considering only final photons, for example, 
    eliminates this problem.
    
    The resulting null zone is consistent with the zero-charge 
    limit of the general null zone conditions.  And in fact it is often useful 
    to think of the final (or initial) multi-photon subset as a massless 
    composite particle. 
    
    We cannot write radiation amplitudes for 
    multi-boson emission in which the zeros are made manifest by a 
    series in $\Delta_{ij}$ factors.  As seen in the examples (\ref{eq:amp})
    and (\ref{eq:gluon}), some terms do not have these 
    factors, yet vanish in the null zone by virtue of their 
    momentum and current dependence.  These terms are again related to the
    currents $J_r$ analyzed above.
    
    We can, however, establish simple forms 
    in the limit where all photons have the same momentum.  This is 
    related to the all-orders solution for an external plane-wave field 
    coming up next.
    \setcounter{equation}{0}

    \section{The More General Result:  An External Field}
    
    It has been pointed out previously~\cite{BK} that 
    multiphoton zeros follow from a decoupling theorem for
    the scattering of a system of particles immersed in an exernal
    electromagnetic plane wave.  In this section we review and elaborate
    upon the details showing this
    connection, and we use one of the examples in
    Sec. II to demonstrate the result.
    
    For a particle with charge $Q$ and mass $m$ coupled
    to an external electromagnetic plane wave $A_\mu=A_\mu (n\cdot x)$, $n^2=0$
    (gauge $n\cdot A=0$), the wave functions for spins $0$, $1/2$, $1$ can all
    be written in the form~\cite{BK,BK2} 
    \begin{equation}
    \Psi (x)=ULT\chi (x)
    \label{eq:psi}\end{equation}
    where $\chi$ is the free solution and $U$,$L$,$T$ are local gauge,
    Lorentz, and displacement transformations, respectively.
    Explicitly, we have for spins $\{0,\,1/2\}$ and a free plane wave,
    \begin{equation}
    \chi_p(x)=e^{-ip\cdot x}\,\{1;\,\omega (p)\}
    \end{equation}
    where $p^2=m^2$, $\not{p}\omega =m\omega$, and
    \begin{equation}\begin{array}{ll}
    L=\{1;1+{Q\over 2n\cdot p}\not{n}\not{A}\} & \\ 
    U(\theta )=e^{i\theta}, & \quad\theta={Q^2\over 2n\cdot p}\int^{n\cdot x}
    dz\, A^2(z) \\
    T(d)=e^{-ip\cdot d}, & \quad d^\mu={Q\over n\cdot p}\int^{n\cdot x} dz\,
    A^\mu (z) 
    \end{array}\label{eq:trans}\end{equation}
    The spin-one results can be found in the references~\cite{BK,BK2}.
    
    Consider initially the
    scattering of a system of particles with no external field.  The tree
    amplitude is
    \begin{equation}
    {\cal T}=\prod_V\prod_I\int dp_I\, D(p_I)V(k)
    \label{eq:tree}\end{equation}
    with internal propagators $D(p_I)$ and vertex factors $V(k)$ [k legs and
    including delta
    functions $\delta (\sum^k p_i)$].
    If we turn on the external
    electromagnetic field $A$, the internal and external legs of the tree
    amplitude (\ref{eq:tree}) are altered according to the Fourier transform of
    (\ref{eq:psi}) changing the $\delta$-functions to 
    \begin{equation}
    \delta_{ext}=\prod_{j=1}^k (ULT)_j \delta (\sum_{i=1}^k p_i)
    \label{eq:delt}\end{equation}
    where it is understood that we replace $n\cdot x$ by $in\cdot\partial/\partial p_j$  
    in the $(ULT)_j$.  Supplementary changes for vertices with derivative
    couplings are discussed in~\cite{BK,BK2}, but in any case (\ref{eq:delt})
    helps us understand the changes in
    particle momenta due to the external field. 
    For a monochromatic external wave,
    $A_\mu =2 Re\, (N\epsilon_\mu e^{-iq\cdot x})$, with frequency $\omega$
    and momentum $q=\omega n$, we see how harmonics arise through the
    identity
    \begin{equation}
    (e^{\pm q\cdot\partial/\partial p})^l\delta (p)=\delta (p\pm lq)
    \end{equation}
    
    To see the generalized radiation zero, we ask that the conditions
    analogous to (\ref{eq:same2}) be satisfied
    \begin{equation}
    {Q_i\over n\cdot p_i}=\mbox{same for all external particles $i$}\label{eq:same5}
    \end{equation}
    The effect of $\partial/\partial p_j$ on the delta function in
    (\ref{eq:delt}) is independent of $j$, implying the various 
    $\theta_j$ and $d_j$ of (\ref{eq:trans}) are also independent of $j$.  From
    charge conservation, Lorentz invariance, and momentum conservation,
    all the phases (group parameters) cancel out:
    \begin{equation}
    \prod_{j=1}^k (ULT)_j=1, \mbox{    in the ``null zone'' defined by
    (\ref{eq:same5})}
    \label{eq:prod}\end{equation}
    We see that the external field effects have disappeared; the field is
    decoupled in the null zone.  Even though it may have been kinematically
    allowed for the particle system to evolve to some final state under the
    influence of the external field, the probability amplitude for that is
    zero!
    
    An expansion order-by-order of (\ref{eq:delt}) in the various charges of the particles is in one-to-one correspondence with the sequence of amplitudes
    for $n$ collinear photons.  For example, attaching $n$ photons with the
    same momentum $q$ (and polarization) to a given leg in all possible ways, remembering the
    seagull graphs for scalars, leads to an exponential form when summed
    over $n$.  In this way, we see the connection between the zeros for an
    $n$-photon amplitude and the decoupling theorem.  When the exponentials
    collapse to unity, every such amplitude is zero.  To generalize to
    photons with different polarizations, we can replace $QA$ by $Q_1A_1 + Q_2A_2 + ...$ in (\ref{eq:trans}) and, as long as the different external fields
    are collinear with respect to their null vectors $n_i$, the forms
    (\ref{eq:psi}) and (\ref{eq:delt}) continue to hold.  The expansion in
    charge produces now the more general $n$-photon amplitudes, with
    independent polarization for each photon.  Again a decoupling theorem
    exists and implies the higher-order radiation zeros
    amplitude-by-amplitude. 
    
    It is satisfying to compare an expansion of (\ref{eq:delt}) with our examples.  Equation (\ref{eq:amp}) in the limits $q_1=q_2=q$ and
    $\epsilon_1=\epsilon_2=\epsilon$ is
    \begin{equation}\begin{array}{rl}
    M & =g\sum_i \left( {Q_i\over p_i\cdot q} -{Q_j\over p_j\cdot q}\right)\, 
    \delta_i p_i\cdot\epsilon^*\,\sum_k\left( {Q_k\over p_k\cdot q}
    -{Q_l\over p_l\cdot q}\right)\,\delta_k p_k\cdot\epsilon^* \\
    & -{1\over 2}g(\epsilon^* )^2\sum_i\left( {Q_i\over p_i\cdot q}
    -{Q_j\over p_j\cdot q}\right)\,\delta_i Q_i
    \end{array}\end{equation}
    This checks perfectly against the second-order term; the first line comes
    from two powers of ``$d$ terms", 
    and the second line corresponds to one power of ``$\theta$ terms", refering to
    the nomenclature in (\ref{eq:trans}).
    
    \setcounter{equation}{0}
    \section{Experiments and Discussion}
    
    In this last section we discuss reactions involving the
    production of
    multiple photons, gluons, or supersymmetric partners thereof, which are  
    well approximated by tree amplitudes, and where
    their tree amplitudes have radiation amplitude zeros (RAZ).  We often
    have in mind the possibility that the zeros may be sensitive to 
    fundamental particle parameters.
    
    There have already been some limits set on the $W$ magnetic moment parameter $\kappa$ from $W\gamma$ 
    and radiative $W$ decay at
    CDF (Fermilab) and UA2 (CERN).  The results~\cite{SLSSS} are
    \begin{equation}\begin{array}{cc}
    -2.4\leq \kappa\leq 3.7 & \qquad\qquad \mbox{(CDF)} \\
    -3.1\leq \kappa\leq 4.2 & \qquad\qquad \mbox{(UA2)} 
    \end{array}\end{equation}
    So far, since the number of events are limited~\cite{Al}, only the total number 
    of events have been used to obtain these limits.  The new run at CDF and
    also D0 will obtain many more events and then one should be able to
    obtain an angular distribution and, hopefully, see the 
    RAZ.  For a related discussion, see~\cite{RS}. Also, 
    the rapidity correlation study by Baur et al.~\cite{Mont} is a new and effective tool in the radiation zero analysis.
    
    Of course the RAZ occurs only if $\kappa_W=1$, or $g_W=\kappa_W+1=2$.  Thus 
    this is a test of the Standard Model (SM) in which $\kappa_W=1$ (plus radiative
    corrections).  Recently Brodsky and Hiller~\cite{BH} have shown that a
    composite particle has in general non-standard magnetic and quadrupole moments.
    However, in the limit of zero radius the moments take their SM values.
    This has been shown for spin 1.  The spin 1/2 case was treated
    earlier~\cite{BD}.  Thus the RAZ in the reactions described previously
    and in what follows 
    constitute a test of the compositeness of the $W$ boson.
    
    Consider first the $2\gamma$ double bremsstrahlung process
    \begin{equation}
    Q_1+Q_2\rightarrow Q_1 +Q_2+\gamma +\gamma
    \end{equation}
    This process is being studied by Ward et al.~\cite{Ward}.  As a check
    of this calculation, one may impose the null-zone conditions described
    in Sec. II, irrespective of whether the zero is physical or not.  The
    differential cross section should then vanish.  If it does not, there is
    an error in the calculation.  Such computational checks are a useful
    feature of the presence of radiation zeros, particularly in the
    higher-order QED and QCD calculations that have become increasingly relevant
    to experimental analysis.  
    
    We come back now to the processes 
    \begin{equation}
    W\rightarrow d\bar{u}\gamma g
    \label{5.2}\end{equation}
    \begin{equation}
    d\bar{u}\rightarrow W\gamma g
    \label{5.3}\end{equation} 
    We have described earlier in Sec. II that these reactions have zeros
    essentially at the same places as the original reactions without the gluons, but
    now with the gluon and photon traveling together.
    These gluon processes could be seen in~\cite{Smith} 
    \begin{eqnarray}
    pp(\bar{p})\rightarrow W\gamma gX\rightarrow (e,\mu)\nu\gamma gX\\
    pp(\bar{p})\rightarrow WX\rightarrow (e,\mu)\nu\gamma gX\nonumber
    \label{5.4}\end{eqnarray}
    where a sharp dip should persist.  These again occur only if $\kappa_W=1$,
    or $g_W=2$ and, therefore, are a test of the SM.  Here we must
    be able to distinguish gluon jets from quark jets and remove the $q\bar{q}$
    background.  One could imagine tagging gluon jets with photons, and
    thereby verifying the consistency of jet identification algorithms.  One
    would look for photons inside gluon jets and find no events when the
    zero conditions are satisfied.  Such an experiment, although 
    difficult would be very interesting.
    
    The difficulty of detecting the photon and jet together has been
    emphasized by Diakonos et al.~\cite{Diak}.  In their recent paper, they
    refer to the neutral particle extension~\cite{BKB} of the radiation zeros,
    noting that the zero arising when the photon and gluon are parallel (and
    at the original RAZ magic angle) 
    is a powerful check on the matrix element calculation.
    Although they are much less sanguine about the experimental consequences,
    there is the possibility that the recent approach by Baur et
    al.~\cite{Mont} may be adapted to the $W$-photon-gluon final state, and
    improve the signal to noise.
    
    One could also look at HERA (DESY) for
    the process~\cite{Bilc,Cout,RLS}
    \begin{equation}
    e^+u\rightarrow e^+u\gamma g
    \end{equation}
    This could be seen in 
    \begin{equation}
    e^+p\rightarrow e^+p\gamma gX 
    \end{equation}
    where a dip should persist.  In the corresponding process 
    \begin{equation}
    e^-p\rightarrow e^-p\gamma g
    \end{equation}
    the zero is washed out~\cite{Cout,RLS}.
    
    In addition, we can consider supersymmetric versions of the radiation zeros~\cite{BK3,DGT}.  For example, in
    \begin{equation}
    \chi^+\rightarrow \chi^0 u\bar{d}
    \end{equation}
    a RAZ occurs in the supersymmetric limit, when
    \begin{equation}
    \tan{\beta} =1
    \end{equation}
    and the masses are equal.  In the context of this paper, it
    is to be noted that adding photons or gluons, or their supersymmetric
    partners, again does not subtract the zeros.
    
    Recently, Ohnemus and Stirling~\cite{OS} considered the process
    \begin{equation}
    pp\rightarrow W^\pm\gamma\gamma X
    \label{5.10}\end{equation}
    which is obtained from the elementary processes 
    \begin{equation}
    q+g\rightarrow W+\gamma+q(\rightarrow\gamma X)
    \end{equation}
    \begin{equation}
    q+\bar{q}\rightarrow W+\gamma+g(\rightarrow\gamma X)
    \label{5.11b}\end{equation}
    \begin{equation}
    q+\bar{q}\rightarrow W+g(\rightarrow\gamma X)+g(\rightarrow\gamma X)
    \label{5.11c}\end{equation}
    \begin{equation}
    q+\bar{q}\rightarrow W+q(\rightarrow\gamma X)+\bar{q}(\rightarrow\gamma X)
    \end{equation}
    \begin{equation}
    q+g\rightarrow W+q(\rightarrow\gamma X)+g(\rightarrow\gamma X)
    \end{equation}
    \begin{equation}
    g+g\rightarrow W+q(\rightarrow\gamma X)+\bar{q}(\rightarrow\gamma X)
    \end{equation}
    These processes were considered as background to the search for the
    Higgs boson via associated production with $W$ bosons.  This process 
    \begin{equation}
    pp\rightarrow W+H(\rightarrow\gamma\gamma ) +X
    \end{equation}
    provides a very clean signature and could be used at the SSC
    or the LHC to find the Higgs.  Process (\ref{5.11b}) 
    has the physical RAZ we have been talking about.  One could hope that a dip persists in
    (\ref{5.10}); we note again the recent work of Baur et al.~\cite{Mont}.  A rough estimate suggests that one could obtain 200
    such events at the SSC.

    {\large {\bf ACKNOWLEDGMENTS}}

    It is a pleasure for MAS to acknowledge the Aspen Center for Physics and the
    theory group at SLAC for their gracious
    hospitality, and he would also like to thank Stanley Brodsky and George Siopsis 
    for valuable discussions. 
    His work was supported by the U. S. Department of Energy under Grant No. 
    DE-FG05-84ER40215.  RWB and MEC are supported by the National Science Foundation 
    and its REU program, and the CWRU industrial problem solving group, and are grateful to Kenneth Kowalski for discussions on 
    these matters through the years.

    \vfill
    \newpage
    {\large\centerline{\bf Figure Captions.}}
    \vskip5ex\noindent
    {\large\bf 1.} Lowest-order diagrams for decay of a scalar particle through a
    single scalar interaction into $n-1$ scalar particles and a photon.
    \newline {\large\bf 2.} Diagrams for a photon attached to a sample tree source
    graph with one internal line.
    \newline {\large\bf 3.} Diagrams for the radiative decay process where 
    quark-antiquark annihilation leads to a W-boson plus a photon and a
    gluon.
    
    \end{document}